\documentclass[12pt,preprint]{aastex}

\newcommand{\wfcfuv}{$U_{\rm 225}$}
\newcommand{\wfcnuv}{$U_{\rm 275}$}
\newcommand{\wfcuv}{$U_{\rm 336}$}
\newcommand{\acsb}{$B_{\rm 435}$}
\newcommand{\acsv}{$V_{\rm 606}$}
\newcommand{\acsi}{$i_{\rm 775}$}
\newcommand{\acsz}{$z_{\rm 850}$}

\newcommand{\wfch}{$H_{\rm 160}$}
\newcommand{\sqmin}{arcmin$^2$}
\newcommand{\uvdrops}{$z\!\simeq\!1\!-\!3$}
\newcommand{\etal}{{et\thinspace al.}}
\newcommand{\Ho}{$H_{\rm 0}$}
\newcommand{\tabref}[1]{Table~\ref{#1}}
\newcommand{\figref}[1]{Figure~\ref{#1}}
\newcommand{\secref}[1]{\S~\ref{#1}}

\begin{document}

\title{UV-dropout Galaxies in the GOODS-South Field from WFC3 Early
  Release Science Observations}

\shorttitle{UV-dropouts in GOODS-S}

\author{
N.~P.~Hathi\altaffilmark{1},  
R.~E.~Ryan~Jr.\altaffilmark{2}, 
S.~H.~Cohen\altaffilmark{3},
H.~Yan\altaffilmark{4}, 
R.~A.~Windhorst\altaffilmark{3}, 
P.~J.~McCarthy\altaffilmark{5}, 
R.~W.~O'Connell\altaffilmark{6},
A.~M.~Koekemoer\altaffilmark{7},
M.~J.~Rutkowski\altaffilmark{3},
B.~Balick\altaffilmark{8},
H.~E.~Bond\altaffilmark{7},
D.~Calzetti\altaffilmark{9},
M.~J.~Disney\altaffilmark{10},
M.~A.~Dopita\altaffilmark{11},
Jay~A.~Frogel\altaffilmark{12},
D.~N.~B.~Hall\altaffilmark{13},
J.~A.~Holtzman\altaffilmark{14},
R.~A.~Kimble\altaffilmark{15},
F.~Paresce\altaffilmark{16},
A.~Saha\altaffilmark{17},
J.~I.~Silk\altaffilmark{18},
J.~T.~Trauger\altaffilmark{19},
A.~R.~Walker\altaffilmark{20},
B.~C.~Whitmore\altaffilmark{7},
and
E.~T.~Young\altaffilmark{21}
}

\altaffiltext{1}{Department of Physics and Astronomy, University of
  California, Riverside, CA 92521}

\altaffiltext{2}{Department of Physics and Astronomy, University of
  California, Davis, CA 92616}

\altaffiltext{3}{School of Earth and Space Exploration, Arizona State
  University, Tempe, AZ 85287-1404}

\altaffiltext{4}{Center for Cosmology and AstroParticle Physics, 
The Ohio State University, Columbus, OH 43210}

\altaffiltext{5}{Observatories of the Carnegie Institute of
  Washington, Pasadena, CA 91101, USA}

\altaffiltext{6}{Department of Astronomy, University of Virginia,
Charlottesville, VA 22904-4325}

\altaffiltext{7}{Space Telescope Science Institute, Baltimore, MD 21218}

\altaffiltext{8}{Department of Astronomy, University of Washington,
Seattle, WA 98195-1580}

\altaffiltext{9}{Department of Astronomy, University of Massachusetts,
Amherst, MA 01003}

\altaffiltext{10}{School of Physics and Astronomy, Cardiff University,
 Cardiff CF24 3AA, United Kingdom}

\altaffiltext{11}{Research School of Astronomy \& Astrophysics,  The
Australian National University, ACT 2611, Australia}

\altaffiltext{12}{Association of Universities for Research in
Astronomy,  Washington, DC 20005}

\altaffiltext{13}{Institute for Astronomy, University of Hawaii,
Honolulu, HI 96822}

\altaffiltext{14}{Department of Astronomy, New Mexico State
University, Las Cruces, NM 88003}

\altaffiltext{15}{NASA--Goddard Space Flight Center, Greenbelt, MD
  20771}

\altaffiltext{16}{Istituto di Astrofisica Spaziale e Fisica Cosmica,
  INAF, Via Gobetti 101, 40129 Bologna, Italy }

\altaffiltext{17}{National Optical Astronomy Observatories, Tucson, AZ
85726-6732}

\altaffiltext{18}{Department of Physics, University of Oxford, Oxford
OX1 3PU, United Kingdom}

\altaffiltext{19}{NASA--Jet Propulsion Laboratory, Pasadena, CA 91109}

\altaffiltext{20}{Cerro Tololo Inter-American Observatory,
La Serena, Chile}

\altaffiltext{21}{NASA--Ames Research Center, Moffett Field, CA 94035}

\email{Nimish.Hathi@ucr.edu}
\shortauthors{Hathi et al}

%--------------------------------------------------

\begin{abstract}

  We combine new high sensitivity ultraviolet (UV) imaging from the
  Wide Field Camera 3 (WFC3) on the Hubble Space Telescope (HST) with
  existing deep HST/Advanced Camera for Surveys (ACS) optical images
  from the Great Observatories Origins Deep Survey (GOODS) program to
  identify UV-dropouts, which are Lyman break galaxy (LBG) candidates
  at \uvdrops. These new HST/WFC3 observations were taken over
  50~\sqmin\ in the GOODS-South field as a part of the Early Release
  Science program. The uniqueness of these new UV data is that they
  are observed in 3 UV/optical (WFC3 UVIS) channel filters (F225W,
  F275W and F336W), which allows us to identify three different sets
  of UV-dropout samples.  We apply Lyman break dropout selection
  criteria to identify F225W-, F275W- and F336W-dropouts, which are
  $z\!\simeq\!1.7$, $2.1$ and $2.7$ LBG candidates, respectively.  We
  use multi-wavelength imaging combined with available spectroscopic
  and photometric redshifts to carefully access the validity of our
  UV-dropout candidates. Our results are as follows: (1) these WFC3
  UVIS filters are very reliable in selecting LBGs with
  $z\!\simeq\!2.0$, which helps to reduce the gap between the well
  studied $z\!\gtrsim\!3$ and $z\!\sim\!0$ regimes; (2) the combined
  number counts with average redshift $z\!\simeq\!2.2$ agrees very
  well with the observed change in the surface densities as a function
  of redshift when compared with the higher redshift LBG samples; and
  (3) the best-fit Schechter function parameters from the rest-frame
  UV luminosity functions at three different redshifts fit very well
  with the evolutionary trend of the characteristic absolute
  magnitude, $M^*$, and the faint-end slope, $\alpha$, as a function
  of redshift.  This is the first study to illustrate the usefulness
  of the WFC3 UVIS channel observations to select $z\!\lesssim\!3$
  LBGs.  The addition of the new WFC3 on the HST has made it possible
  to uniformly select LBGs from $z\!\simeq\!1$ to $z\!\simeq\!9$, and
  significantly enhance our understanding of these galaxies using HST
  sensitivity and resolution.

\end{abstract}


%--------------------------------------------------

\section{Introduction}\label{introduction}

The Lyman break `dropout' technique was first applied to select Lyman
break galaxies (LBGs) at $z\!\simeq\!3$ \citep{guha90,stei96,stei99},
and since then it has been extensively used to select LBG candidates
at $z\!\simeq\!3\!-\!9$
\citep[e.g.,][]{sawi06,bouw07,redd08,rafe09,oesc10,bunk10,yan10}.
This dropout technique has generated large samples of star-bursting
galaxy candidates at $z\!\simeq\!3\!-\!9$, but there is only one major
study \citep{ly09} that investigates LBGs at \uvdrops\ based on
dropout selection criteria.  The primary reason for this is that we
need highly sensitive space-based cameras to observe the mid- to
near-ultraviolet (UV) wavelengths required to select LBGs at
\uvdrops. The new Wide Field Camera 3 (WFC3) on the refurbished Hubble
Space Telescope (HST) with its superior sensitivity --- compared to
the Wide-Field Planetary Camera 2 (WFPC2) or the Galaxy Evolution
Explorer (GALEX) --- and filters below the atmospheric cut-off
wavelength (e.g., F225W and F275W), allows us to photometrically
identify and study lower redshift (\uvdrops) LBGs. The improved
sensitivity/depth allows us to probe the lower luminosity systems at
these redshifts. 

There are two important reasons to understand these
LBGs. First, to study the star formation properties of these LBGs,
because they are at redshifts corresponding to the peak epoch of the
global star formation rate
\citep[e.g.,][]{ly09,bouw10a,bouw10b,yan10}, and, secondly, they are
likely lower redshift analogs of the high redshift LBGs --- because of
the similar dropout selection at all redshifts --- whose understanding
will help shed light on the process of reionization in the early
Universe \citep[e.g.,][]{labb10,star10,yan10}. The major advantage of
identifying and studying various properties --- including star
formation properties --- of lower redshift LBGs is that these LBGs
can be investigated in rest-frame UV \emph{as well as} rest-frame
optical filters.  The high redshift LBGs have very little information
on their rest-frame \emph{optical} properties, so a detailed
understanding of lower redshift LBGs is very important to get insight
into the physical and morphological nature of high redshift LBGs.

The new UV observations of the WFC3 Science Oversight Committee (SOC)
Early Release Science extragalactic program (PID: 11359, PI:
O'Connell; hereafter ``ERS2''), covers approximately 50~\sqmin\ in the
north-western part of the Great Observatories Origins Deep Survey
\citep[GOODS;][]{giav04a} South field. Here we use the high
sensitivity of the new WFC3 UVIS channel data, along with existing deep
optical data obtained with the Advanced Camera for Surveys (ACS) as
part of the GOODS program, to search for LBG candidates at \uvdrops.
We use dropout color selection criteria based on color-color plots,
obtained with the WFC3 UVIS and ACS filters to find three unique sets of UV
dropouts --- F225W-dropouts, F275W-dropouts and F336W-dropouts
--- which are LBG candidates at $z\!\simeq\!1.7$, $2.1$ and $2.7$,
respectively (as shown in \figref{fig:lybreak}).

This paper is organized as follows: In \secref{data} we summarize the
WFC3 ERS2 observations, and in \secref{sample} we discuss the
selection, and in \secref{reliable} the reliability of our color
selected \uvdrops\ LBG sample.  In \secref{results} we discuss the
data analysis, which includes measuring their number counts and
surface density (\secref{ncounts}), and compare these with other
surveys at higher redshifts, and estimate rest-frame UV luminosity
functions (\secref{lfs}) for these samples.  In \secref{conclusion} we
conclude with a summary of our results.

In the remaining sections of this paper we refer to the
HST/WFC3 F225W, F275W, F336W, filters as \wfcfuv, \wfcnuv,
\wfcuv, and HST/ACS F435W, F606W, F775W, F850LP filters as
\acsb, \acsv, \acsi, \acsz, respectively, for convenience.  We assume
a \emph{Wilkinson Microwave Anisotropy Probe} (WMAP) cosmology with
$\Omega_m$=0.274, $\Omega_{\Lambda}$=0.726 and
\Ho=70.5~km~s$^{-1}$~Mpc$^{-1}$, in accord with the 5 year WMAP
estimates of \citet{koma09}.  This corresponds to a look-back time of
10.37~Gyr at $z\!\simeq\!2$.  Magnitudes are given in the AB$_{\nu}$
system \citep{oke83}.

%--------------------------------------------------

\section{Observations}\label{data}

\begin{figure}
\epsscale{0.8}
\plotone{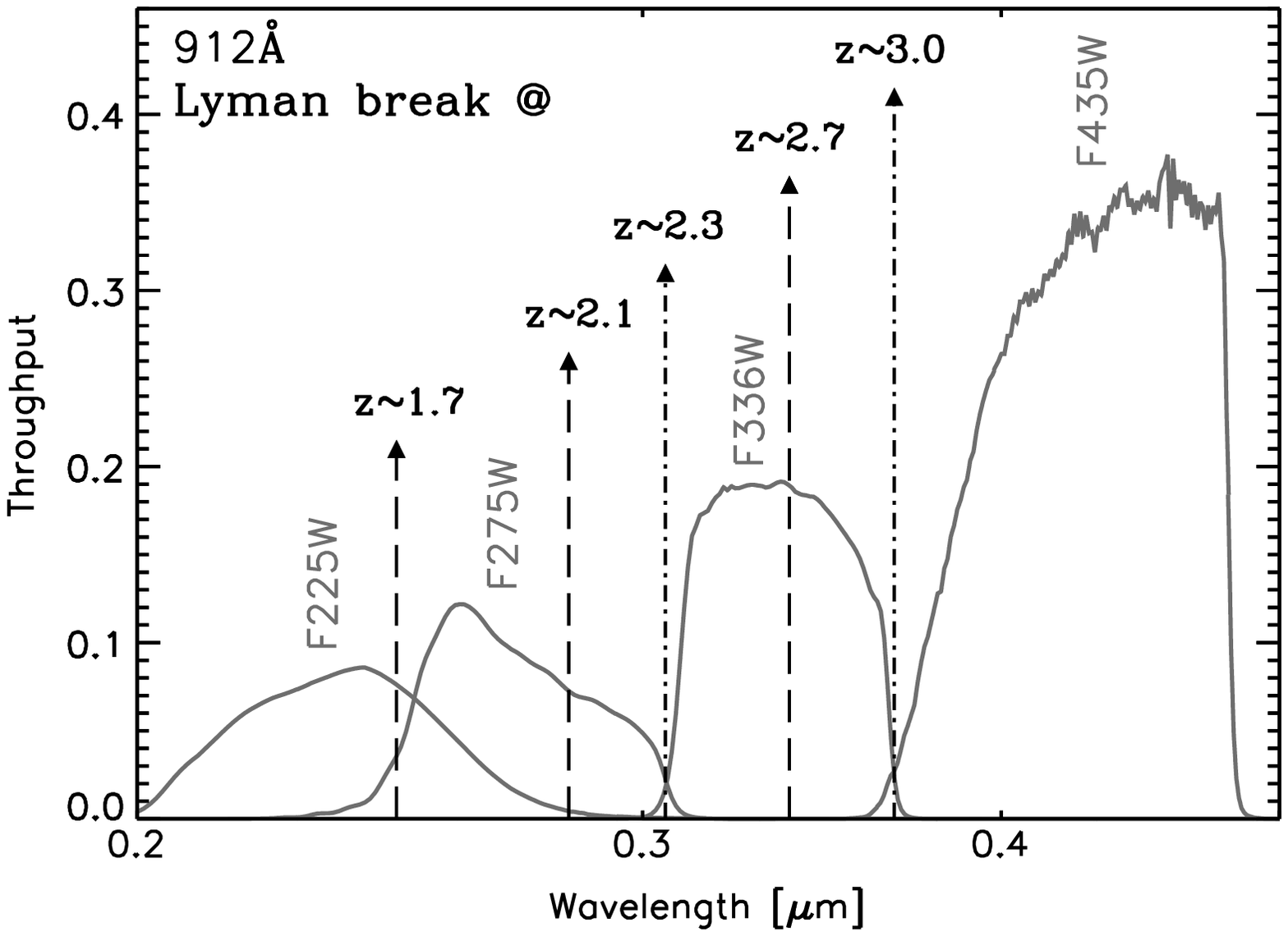}
\caption{Rest-Frame 912~\AA\ Lyman break at different redshifts is shown
  with filter transmission curves of three WFC3 UVIS filters and the
  \acsb-band ACS optical filter. It is evident that \uvdrops\ LBG
  candidates can be efficiently selected using these three WFC3 UV
  filters.}\label{fig:lybreak}
\end{figure}

The WFC3 ERS2 observations were done in both the UVIS (with a FOV of
7.30 \sqmin) and the IR (with a FOV of 4.65 \sqmin) channels. Details
of these observations are described in \citet{wind10}. Here we briefly
summarize the UV imaging observations.  The WFC3 ERS2 UV observations
were done in three broad-band filters \wfcfuv, \wfcnuv\ and \wfcuv,
whose total throughput curves are shown in \figref{fig:lybreak}. The
ERS2 field covers the north-western  $\sim$50~\sqmin\ of the
GOODS-South field, and was observed in 8 pointings with a 2$\times$4
grid pattern during September-October 2009. The \wfcfuv\ and \wfcnuv\
filters were observed for 2 orbits per pointing, while the \wfcuv\
filter was observed for 1 orbit per pointing, for a total of 40 orbits
over the full ERS2 field. The raw images were processed through the
\texttt{CALWF3} task (using the latest version as of December 1, 2009)
included in the STSDAS package (version 3.11), and the latest
reference files from the STScI. The flat-fielded images were then
aligned and drizzled using \texttt{MULTIDRIZZLE} \citep{koek02} onto
the same grid as the GOODS-South
v2.0\footnote{http://archive.stsci.edu/pub/hlsp/goods/v2/} ACS data,
which were rebinned to a pixel size of 0.09\arcsec. The final UV image
mosaics have a pixel scale of 0.09\arcsec\ --- to match the WFC3 IR
image mosaics --- in all filters, and cover $\sim$50~\sqmin\ area in
the GOODS-South field.

The combination of the three WFC3 UV filters and the four ACS optical
filters provide excellent capability of selecting galaxies at
\uvdrops, using the dropout technique to detect the Lyman-break
signature that occurs at a rest-frame wavelength of 912~\AA\
\citep{mada95}. \figref{fig:lybreak} shows the locations of the
rest-frame 912~\AA\ Lyman break at various redshifts. It is clear that
three WFC3 UVIS filters, along with the ACS \acsb- and \acsv-bands are
very useful in identifying LBG candidates at \uvdrops.  We performed
matched-aperture photometry by running the \texttt{SExtractor}
\citep{bert96} algorithm in the dual-image mode with the corresponding RMS
maps. The RMS maps were derived from the \texttt{MULTIDRIZZLE}
generated weight maps, following the procedure discussed in
\citet{dick04}. We have measured the 10-$\sigma$ point source
detection limits in a 0.2\arcsec\ aperture as 26.0, 26.1 and 25.7~AB-mag
in \wfcfuv, \wfcnuv\ and \wfcuv, respectively. We constructed three
separate catalogs by using the three separate images (\wfcnuv, \wfcuv,
\acsb) as detection images. These catalogs are referred as the
\wfcnuv-based, \wfcuv-based and \acsb-based catalogs. We used the WFC3
in-flight photometric zeropoints \citep[24.06, 24.14, 24.64 AB-mag for
\wfcfuv, \wfcnuv\ \& \wfcuv ;][]{kali09} obtained from STScI
website\footnote{zeropoints were made public in September 2009:
  http://www.stsci.edu/hst/wfc3/phot\_zp\_lbn/}.

%--------------------------------------------------

\subsection{Color Selection}\label{sample}

\begin{figure}
\begin{center}
\includegraphics[scale=0.725,angle=90]{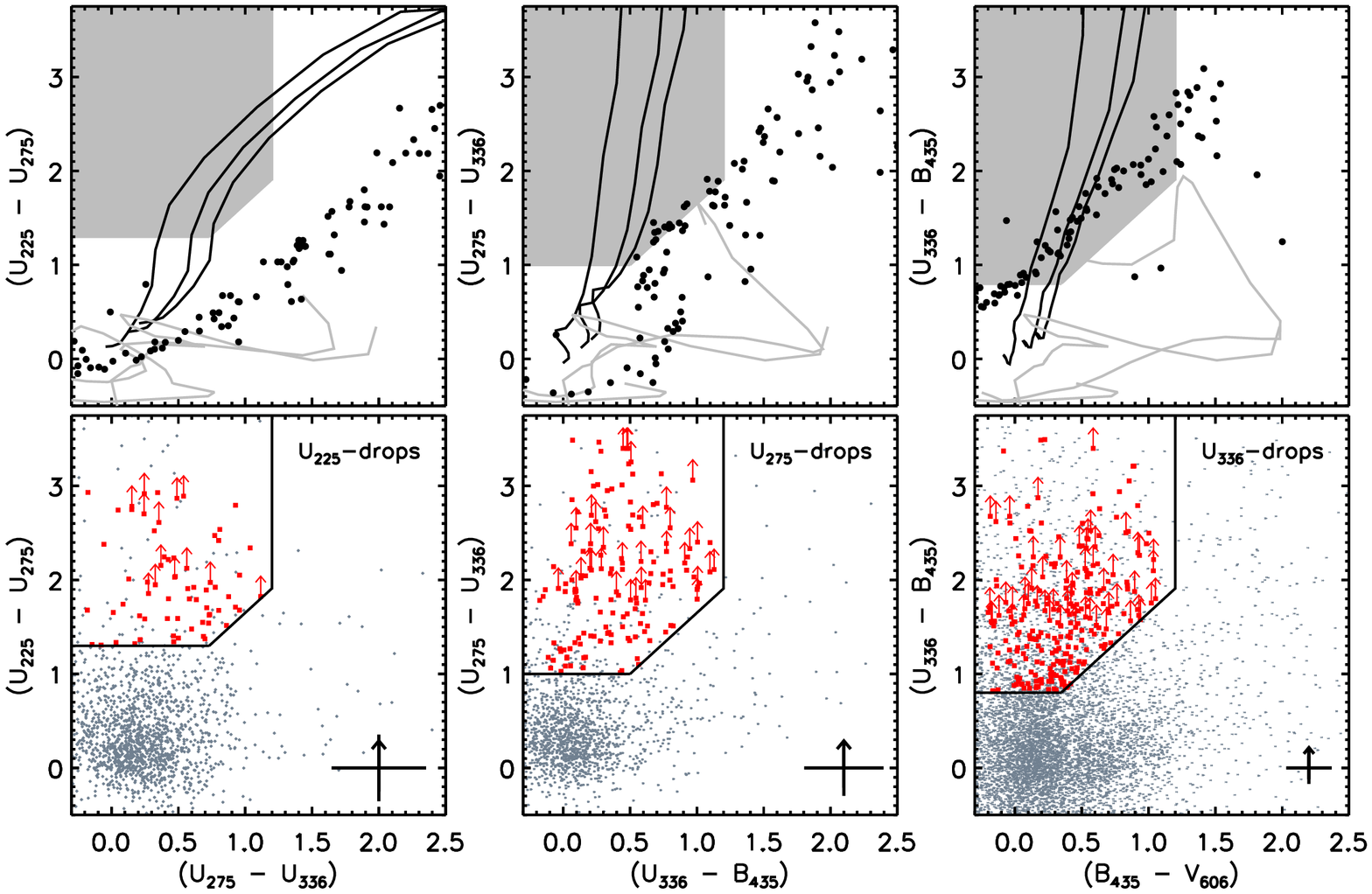}	
\caption{The top three panels show color selection region (gray shaded
  region) obtained using BC03 star-forming galaxy models with
  $E(B-V)=0, 0.15, 0.30$~mag (solid black lines), expected colors of
  stars (black dots) from \citet{pick98}, and tracks of low-redshift
  ellipticals (gray lines) from \citet{kinn96} and
  \citet{cole80}. Though stars clearly land in the selection region,
  these are easily removed by simple morphological criteria, as in
  \citet{wind10}. [Bottom-Left] shows the color-color plot with the
  \wfcfuv-dropout selection region, [Bottom-Middle] shows the
  selection of the \wfcnuv-dropouts and [Bottom-Right] shows the
  selection of the \wfcuv-dropouts. The gray data points in the bottom
  panels are \emph{all} objects in the catalog.  Average uncertainties
  in the color measurements are shown as the error bar in the lower
  right corner. Red points indicate the selected dropouts, while gray
  data points in the selected region were excluded by other criterion
  as given in \secref{sample}.}\label{fig:clr-clr}
\end{center}
\end{figure}

Our initial selection of UV dropouts is based on dropout color
criteria obtained from the stellar population models of
\citet[BC03]{bruz03}. The top three panels of \figref{fig:clr-clr}
shows the BC03 star-forming galaxy models for three dropout samples
with $E(B-V)=0, 0.15, 0.30$~mag (solid black lines), expected colors of
stars (black dots) from \citet{pick98}, and tracks of low-redshift
ellipticals (gray lines) from \citet{kinn96} and \citet{cole80}. The
gray shaded region in top panels is our selection region. Though
Galactic stars clearly land in the selection region in
\figref{fig:clr-clr}, these are easily removed by simple morphological
criteria, as in \citet{wind10}. We have applied the \citet{mada95}
prescription to estimate IGM attenuation for proper comparison with
other studies.  We also checked how BC03 tracks are affected by
applying the Madau prescription only to galaxies at $z\!>\!2.5$
(i.e. no IGM attenuation below $z\!<\!2.5$), and only to galaxies at
$z\!>\!1.0$, to see the effects of fluctuations in IGM attenuation.  We
find that our adopted selection criteria will still be able to pick-up
star-forming galaxies with $E(B-V)<0.3$. The selection criteria
adopted here are similar to the criteria used to identify LBG
candidates at $z\!\simeq\!3\!-\!8$
\citep[e.g.,][]{stei96,giav04b,bouw10a,bouw10b,yan10}.  We use
\wfcnuv-based catalogs to select \wfcfuv-dropouts using (\wfcfuv\ --
\wfcnuv) vs. (\wfcnuv\ -- \wfcuv) color-color space, as shown in the
bottom leftmost panel of \figref{fig:clr-clr}. For \wfcfuv-band
dropouts, we require:
\begin{displaymath}
      \left\{ \begin{array} {ll} (U_{\rm 225}-U_{\rm 275}) > 1.3 \:
        {\rm mag} \: \hbox{and} \: U_{\rm 275} \le 26.5 \:\hbox{mag}
        \\ \hbox{and}\: (U_{\rm 275}-U_{\rm 336}) < 1.2 \:\hbox{mag}
        \\ \hbox{and}\: (U_{\rm 275}-U_{\rm 336}) > -0.2 \:\hbox{mag}
        \\ \hbox{and}\: (U_{\rm 225}-U_{\rm 275}) > 0.35 + [1.3 \times
          (U_{\rm 275}-U_{\rm 336})] \:\hbox{mag} \\ \hbox{and}\:
        (U_{\rm 336}-B_{\rm 435}) > -0.5, \: [S/N(U_{\rm 275})] > 3,\:
           [S/N(U_{\rm 225})] < 3
                \end{array} \right.
\end{displaymath} 
Here, the $S/N$ is defined as 1.0857 divided by the
\texttt{SExtractor} error in the total magnitude.  The
\texttt{SExtractor} magnitude uncertainties are estimated from
carefully constructed RMS maps that account for correlated pixel noise
and hence, gives better estimate of the noise.  Details of the WFC3
data reduction process are given in \citet{wind10}.  We require
(\wfcfuv\ -- \wfcnuv) $>$ 1.3 mag, which is redder than what we have
applied for other dropouts, because there is no bluer filter available
than \wfcfuv\ to confirm that these dropouts are undetected at
wavelengths bluer than rest-frame 912~\AA. We have also applied the
additional criterion (\wfcuv\ -- \acsb) $>$ --0.5 mag to eliminate the
possibility of selecting spurious candidates, since this color-color
space is based only on three UV filters, and it is required that LBG
candidates be detected and are bright enough in the \acsb-band.  The
\acsb-band is about $\sim$2--2.5 mag deeper than the \wfcuv-band, so a
simple $S/N$ (\acsb) $>$ 3 cut cannot be used, because we would still
pick-up very faint objects in \acsb-band. There are 106 objects inside
the selection region. We find a total of 70 LBG candidates
(\wfcfuv-dropouts) based on these selection criteria. After visually
checking each candidate using the 10-band (3 WFC3 UV, 4 ACS optical
and 3 WFC3 IR) HST imaging from the ERS2 GOODS observations, we
eliminated 4 candidates from our sample as spurious (due to their
closeness to a bright object or a probable faint stellar diffraction
spike). This examination leaves us with 66 \wfcfuv-dropouts.

Similarly, we use \wfcuv-based catalogs to select \wfcnuv-dropouts
using (\wfcnuv\ -- \wfcuv) vs. (\wfcuv\ -- \acsb) color-color space as
shown in the bottom middle panel of \figref{fig:clr-clr}. For \wfcnuv-band
dropouts, we require:
\begin{displaymath}
      \left\{ \begin{array} {ll} (U_{\rm 275}-U_{\rm 336}) > 1.0 \:
        {\rm mag} \: \hbox{and} \: U_{\rm 336} \le 26.5 \:\hbox{mag}
        \\ \hbox{and}\: (U_{\rm 336}-B_{\rm 435}) < 1.2 \:\hbox{mag}
        \\ \hbox{and}\: (U_{\rm 336}-B_{\rm 435}) > -0.2 \:\hbox{mag}
        \\ \hbox{and}\: (U_{\rm 275}-U_{\rm 336}) > 0.35 + [1.3 \times
          (U_{\rm 336}-B_{\rm 435})] \:\hbox{mag} \\ \hbox{and}\:
           [S/N(U_{\rm 336})] > 3,\: [S/N(U_{\rm 275})] < 3, \:
           [S/N(U_{\rm 225})] < 1
                \end{array} \right.
\end{displaymath} 
There are 223 objects inside the selection region. We find a total of
153 LBG candidates (\wfcnuv-dropouts) based on these selection
criteria. After visual examination, we eliminated 2 candidates from
our sample as spurious, because of their proximity to a
brighter object. Therefore, our core sample contains 151
\wfcnuv-dropouts.

Finally, we use \acsb-based catalogs to select \wfcuv-dropouts using
(\wfcuv\ -- \acsb) vs. (\acsb\ -- \acsv) color-color space as shown in
the bottom rightmost panel of \figref{fig:clr-clr}. For \wfcuv-band dropouts,
the following color selection was applied:
\begin{displaymath}
      \left\{ \begin{array} {ll} (U_{\rm 336}-B_{\rm 435}) > 0.8 \:
        {\rm mag} \: \hbox{and} \: B_{\rm 435} \le 26.5 \:\hbox{mag}
        \\ \hbox{and}\: (B_{\rm 435}-V_{\rm 606}) < 1.2 \:\hbox{mag}
        \\ \hbox{and}\: (B_{\rm 435}-V_{\rm 606}) > -0.2 \:\hbox{mag}
        \\ \hbox{and}\: (U_{\rm 336}-B_{\rm 435}) > 0.35 + [1.3 \times
          (B_{\rm 435}-V_{\rm 606})] \:\hbox{mag} \\ \hbox{and}\:
           [S/N(B_{\rm 435})] > 3,\: [S/N(U_{\rm 336})] < 3,\:
           [S/N(U_{\rm 275})] < 1, \: [S/N(U_{\rm 225})] < 1
                \end{array} \right.
\end{displaymath}
We require (\wfcuv\ -- \acsb) $>$ 0.8 mag, which is bluer than what we
have applied for other dropouts, because we have two bands bluer than
\wfcuv\ to confirm that these dropouts are not detected ($S/N <1$) at
wavelengths bluer than rest-frame 912~\AA. There are 1156 objects
inside the selection region. We find a total of 260 LBG candidates
(\wfcuv-dropouts) based on the above mentioned selection
criteria. After visually checking each candidate, we eliminated 4
spurious candidates (same reason as discussed before for other dropout
candidates). The final sample consists of 256 \wfcuv-dropouts.

The gray data points in the bottom panels of \figref{fig:clr-clr}
shows \emph{all} the sources in our catalogs that are not selected as
LBG candidates. Those that fall within our selection regions were
excluded from our candidate samples mainly because of our low $S/N$
cuts ($<1$) in the bluer bands and the hard magnitude limit
($<26.5$~mag) in the selection band.  The primary reason for this is
the varying depth of these filters. The \acsb\ is much deeper than
\wfcuv, so when we select \wfcuv-dropouts with \acsb\ $<26.5$~mag, we
are not selecting galaxies fainter than 26.5~mag in \acsb. We cannot
conclusively say whether these faint galaxies are LBG candidates or
not because of the shallower \wfcuv\ images.  Similarly, \wfcfuv\ and
\wfcnuv\ are slightly deeper than \wfcuv, so while selecting the
\wfcnuv- or \wfcuv-dropouts, our selection criteria of $S/N<1$ in the
bluer bands still keeps some faint objects (with $S/N>1$) in the
selection region which are not selected as dropouts.  Therefore, the
varying depth between filters, combined with our magnitude and $S/N$
cuts, are responsible for objects in the selection region that are not
selected as LBG candidates. Overall, our selection is conservative,
because of these constraints.

There are a few compact objects in our selected samples, but when we
visually check these objects in the 10-band HST imaging, and compare
with the more robustly selected star catalog of \citet{wind10}, we
cannot confirm any stars. \citet{wind10} gives a detailed discussion
of the star-galaxy separation procedure used for the ERS2 data, and
confirms that within our sample magnitude range (24--26.5~mag) there
are practically no stars in the UV bands, and very few in the
B-band. There could be a weak AGN in some of these galaxies, which we
will investigate in our future paper on stellar populations and
spectral analysis of these LBGs.

\begin{figure}
\begin{center}
\includegraphics[scale=0.8]{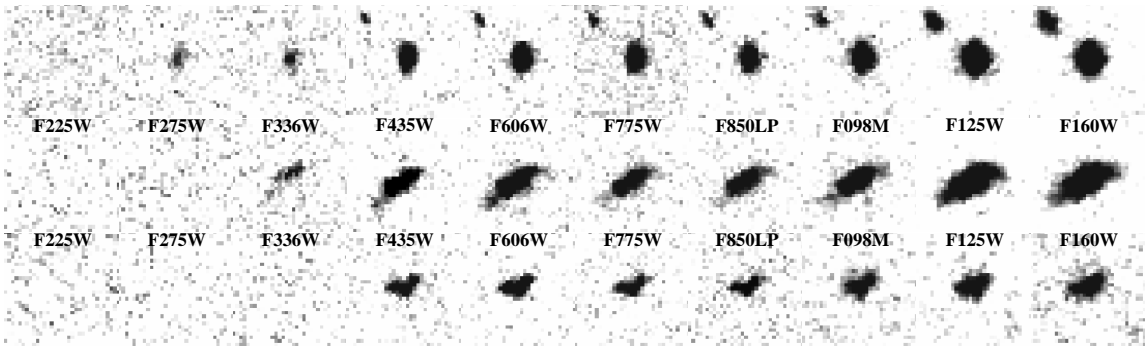}	
\caption{Three examples of color selected UV-dropouts with
  spectroscopic redshifts are shown here in the 10-band HST imaging from
  the ERS2 data. The object at the top is a \wfcfuv-dropout (\acsv\
  $\sim$ 24.5 mag) with a spectroscopic redshift of $z\!\simeq\!1.61$,
  the object in the middle is a \wfcnuv-dropout (\acsv\ $\sim$ 23.9
  mag) with $z\!\simeq\!2.04$, and the object at the bottom is a
  \wfcuv-dropout (\acsv\ $\sim$ 24.7 mag) with $z\!\simeq\!2.69$. Each
  stamp is 3\arcsec\ on a side, has North up, and has a pixel scale of
  0.09\arcsec/pixel.}\label{fig:image}
\end{center}
\end{figure}

\figref{fig:image} shows three example images of color selected
UV-dropouts whose redshifts are confirmed by ground-based spectroscopy
(see next section). These examples are shown here in the 10-band HST
imaging obtained from ERS2 observations.  The final sample consists of
66 \wfcfuv-, 151 \wfcnuv- and 256 \wfcuv-dropouts. There is one object
overlapping between \wfcfuv- and \wfcnuv-dropout samples, while seven
objects are in common between \wfcnuv- and \wfcuv-dropout samples.

\subsection{Reliability of Color Selection}\label{reliable} 

In order to test reliability of our color selection, we compare our
dropout samples with spectroscopic redshifts from the Very Large
Telescope \citep[VLT; e.g.,][]{graz06,wuyt08,vanz08,bale10} and with
the 10-band (3 WFC3 UV, 4 ACS optical and 3 WFC3 IR) photometric
redshifts obtained from our ERS2 observations (Cohen~et~al.~2010,
in~prep). When we match our dropout catalogs with these redshift
catalogs, we find that $\sim$80\% of our dropouts have photometric
redshifts, but only $\sim$30\% have spectroscopic redshifts. Though
photometric redshifts are from the same ERS2 dataset, we don't have
100\% matching catalogs.  The ERS2 photometric redshifts are based on
the \wfch-band selected catalogs, and the WFC3 IR channel covers a
smaller area than the WFC3 UVIS or ACS/WFC cameras, which were used
here to identify these UV dropouts. The most likely reason for the low
number of spectroscopic confirmations is the `redshift desert'. The
galaxies in this redshift range ($1\!\lesssim\!z\!\lesssim\!3$) are
difficult to identify via ground-based spectroscopy, because of the
lack of strong features in 4500-9000~\AA\ range, where most
spectrographs on large telescopes are optimized.

\begin{figure}
\begin{center}	
\includegraphics[scale=0.725,angle=90]{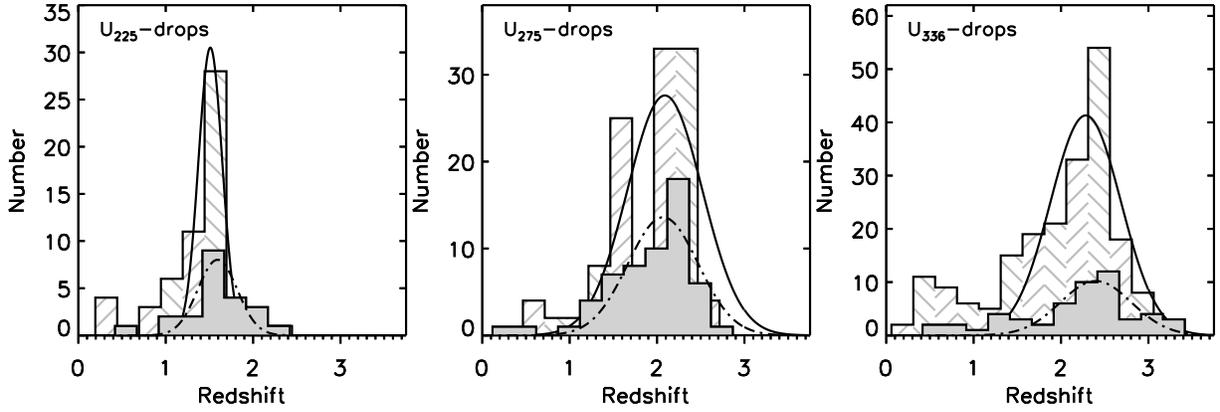}
\caption{ The hashed histogram (solid gray) and solid (dot-dash) curve
  shows the distribution and the Gaussian fit, respectively, to our
  LBG candidates with photometric (spectroscopic) redshifts.  [Left]
  shows distribution of 55 (22) \wfcfuv-dropouts with photometric
  (spectroscopic) redshifts. The average redshifts are
  $<\!z_{ph}\!>$=1.51$\pm$0.13 and $<\!z_{sp}\!>$=1.59$\pm$0.22,
  [Middle] shows the distribution of 117 (57) \wfcnuv-dropouts with
  photometric (spectroscopic) redshifts. The average redshifts are
  $<\!z_{ph}\!>$=2.09$\pm$0.42 and $<\!z_{sp}\!>$=2.07$\pm$0.40, and
  [Right] shows the distribution of 203 (52) \wfcuv-dropouts with
  photometric (spectroscopic) redshifts. The average redshifts are
  $<\!z_{ph}\!>$=2.28$\pm$0.40 and
  $<\!z_{sp}\!>$=2.40$\pm$0.40.}\label{fig:redshifts}
\end{center}
\end{figure}

\figref{fig:redshifts} shows the redshift distributions of the dropout
LBG candidates --- three dropout samples are shown in three separate
panels --- with spectroscopic and photometric redshifts.  The hashed
(solid gray) histogram and solid (dot-dash) curve shows the
distribution and the best Gaussian fit to the number of LBG candidates
with photometric (spectroscopic) redshifts. \tabref{tab:redshifts}
lists the number of dropouts with spectroscopic/photometric redshifts
and their average redshifts obtained from the distributions in
\figref{fig:redshifts}.

\figref{fig:redshifts} shows that based on these available redshifts,
our dropout selection is very reliable, and that WFC3 UV filters provide a
very efficient way to select LBGs at \uvdrops. Spectroscopic redshifts
are only available for $\sim$30\% of our dropout galaxies, and have
--- on average --- $\sim$6\% outliers or low-$z$ interlopers
($\sim$5\% for \wfcfuv-, $\sim$5\% for \wfcnuv- and $\sim$9\% for
\wfcuv-dropouts). The outliers are defined as any object at
$z\!\lesssim\!1$. The ERS2 photometric redshifts are available for
$\sim$80\% of our dropouts and have --- on average --- $\sim$12\%
outliers with the most ($\sim$15\%) amongst the \wfcfuv-dropouts, as
expected because of the lack of any available WFC3 bands bluer than
\wfcfuv. The comparison of our color selected dropouts with
spectroscopic and photometric redshifts show that the fraction of
outliers in both cases are comparable or better than the fraction of
outliers in spectroscopic follow-up surveys of star-forming galaxies
and dropout selected LBGs at $z\!\simeq\!1.5\!-\!3.4$, which is about
$\sim$5-15\% \citep{stei03,stei04,redd08,ly09}.  A similar comparison
with publicly available photometric redshifts
\citep[e.g.,][]{wuyt08,sant09} shows that they have a higher
percentage of outliers ($\sim$17\%), because these large surveys have mostly
shallow ground-based near-UV data which cannot go bluer than
$\sim$3000~\AA, because of the atmospheric cut-off. Therefore,
space-based WFC3 UV data are essential to get accurate photometric
redshifts for these lower redshift galaxies (\uvdrops).

The average spectroscopic or photometric redshift for \wfcuv-dropouts
from \figref{fig:redshifts} is $z\!\simeq\!2.4$, but based on the
location of the Lyman break (\figref{fig:lybreak}), the average
redshift for \wfcuv-dropouts should be about $z\!\sim\!3.0$.  We have
identified small number of brighter ($\la 26.0$~mag) LBG candidates at
$z\!\gtrsim\!2.8$, but we are missing a significant number of fainter
($> 26.0$~mag) LBG candidates at these redshifts. The main reason is
that we require redder colors (\wfcuv\ -- \acsb\ $\gtrsim$ 2.0~mag) to
select higher redshift galaxies in this dropout sample, because the
Ly$\alpha$ forest absorption at 912-1216~\AA\ begins to increasingly
affect the \wfcuv\ band at these redshifts.  This implies that we need
the \wfcuv\ images to be $\sim$1--1.5 mag deeper to consistently
select all dropouts at $z\!\ga\!2.8$, improve the photometric redshift
distribution, and lower the number of outliers for this
sample. Therefore, we have a relatively smaller number of LBG
candidates (within our magnitude limit) at $z\!\gtrsim\!2.8$ in the
\wfcuv-dropouts sample.

%--------------------------------------------------

\section{Results and Discussion}\label{results}
\subsection{Number Counts}\label{ncounts}

\begin{figure}
\epsscale{0.50}
\plotone{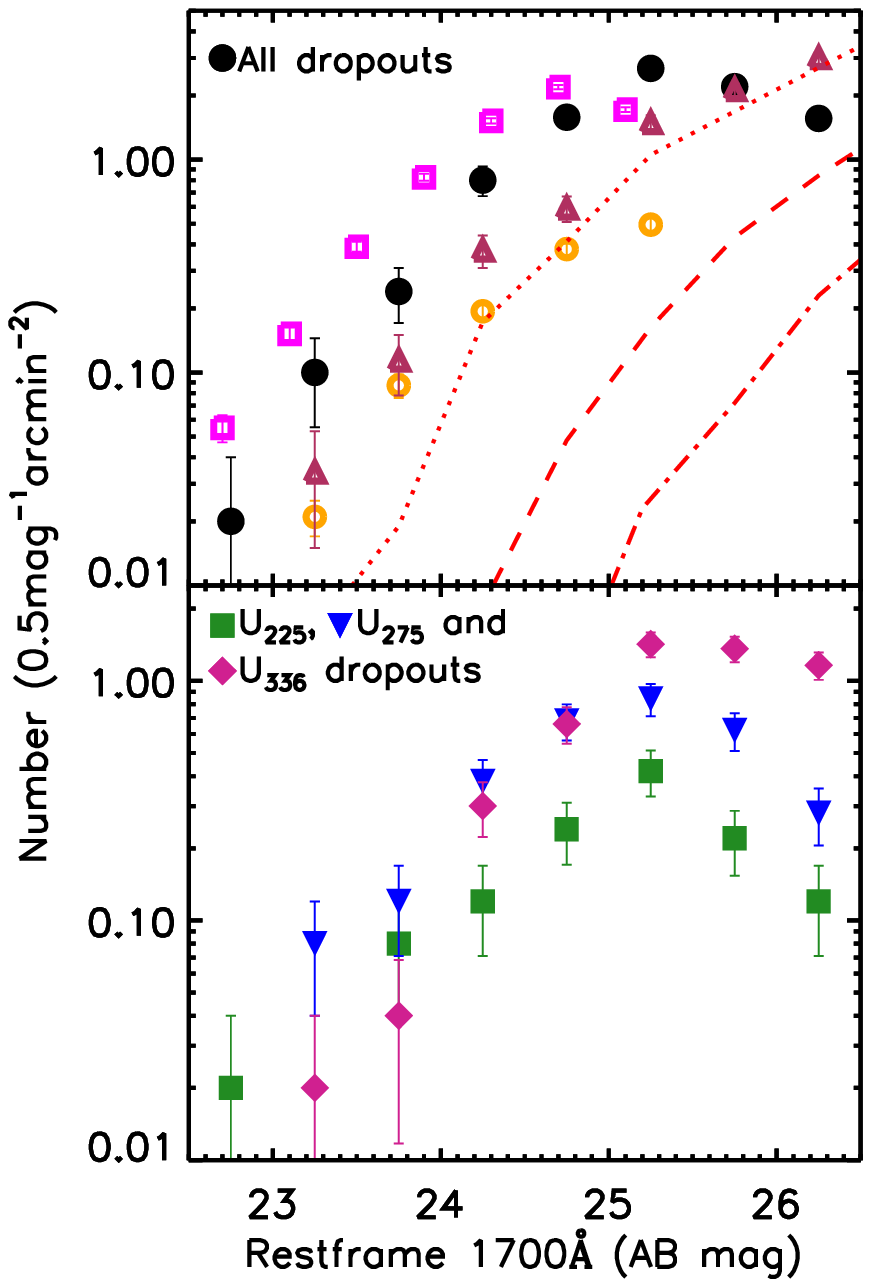}
\caption{[Top] Rest-frame 1700~\AA\ number counts (i.e., number per
  arcmin$^2$ per 0.5 mag bin) of all dropouts ($z\!\simeq\!1\!-\!3$)
  in our sample shown by black solid circles, along with the studies of
  \citet[$z\!\simeq\!3$, orange open circles]{stei99},
  \citet[$z\!\simeq\!3$, maroon open triangles]{noni09} and
  \citet[$z\!\simeq\!2.2$, magenta open squares]{ly09}. The red lines
  are the number counts from \citet{bouw07} for LBGs at
  $z\!\simeq\!4$ (dotted), $z\!\simeq\!5$ (dashed) and $z\!\simeq\!6$
  (dot-dash). [Bottom] The bottom panel shows our number counts for
  each dropout sample (\wfcfuv, \wfcnuv, \wfcuv). The vertical error
  bars in our data are 1$\sigma$ Poisson
  uncertainties.}\label{fig:ncounts}
\end{figure}

The observed raw number counts of LBG candidates at \uvdrops\ at a
rest-frame wavelength of 1700~\AA\ are shown in
\figref{fig:ncounts}. When we combine all three dropout samples, the
average photometric redshift is $z\!\simeq\!2.2$. For proper
comparison, these number counts are \emph{not} corrected for
incompleteness or cosmic variance, and therefore, the counts start to
drop at fainter magnitudes ($\gtrsim26.0$~mag).  \figref{fig:ncounts}
(top panel) shows number counts (in number per arcmin$^2$ per 0.5 mag
bin) of \emph{all} dropouts (\uvdrops) in our sample compared with
other ground-based and space-based LBG surveys
\citep{stei99,noni09,ly09} at $z\!\simeq\!2\!-\!3$. We have also
plotted $z\!\simeq\!4\!-\!6$ number counts from \citet{bouw07} to show
the change in the surface densities (number per arcmin$^2$) as a
function of redshift.

\citet[$z\!\simeq\!3$]{stei99} used ground-based imaging in $\sim$14
fields, with each field observed for many kilo-seconds (ks), followed
by ground-based spectroscopy to confirm many of their color selected
candidates. The \citet{stei99} selection was based on LBG color
criteria down to AB$\sim$25~mag. \citet{noni09} observed the
GOODS-South field using VLT/VIMOS to get deep $U$-band imaging
(AB$\sim$27~mag). Their number counts for LBG candidates at
$z\!\simeq\!3$ shown in \figref{fig:ncounts} come from the deepest
part of the VIMOS field, which covers $\sim$88~\sqmin\ with exposure
time of $\sim$20~hours (72~ks). On the other hand, \citet{ly09}
observed the Subaru Deep Field \citep{kash04} using deep ($>$100~ks)
near-UV imaging from the space-based GALEX observations (with
$\sim$5\arcsec\ FWHM resolution) to select LBG candidates at
$z\!\simeq\!2.2$ down to AB$\sim$25~mag, and used ground-based
spectroscopy to confirm many LBGs at $z\!\simeq\!2.2$.

From \figref{fig:ncounts} (top panel) we note three major
points. First, there is only one space-based --- GALEX --- LBG survey at
$z\!\simeq\!2.2$ \citep{ly09}, which clearly shows that the WFC3 UV
observations --- with better sensitivity and resolution --- can play a
vital role in identifying LBGs at $z\!\lesssim\!2.5$.  Secondly, all
surveys mentioned above use deep UV imaging with $\gtrsim70$~ks
exposures, while our WFC3 UV observations are only $\lesssim5$~ks (1 to
2 orbits), and still we find that our observations are
$\sim$0.5--1.0~mag deeper compared to some of these surveys.  Finally,
our numbers agree very well with the decreasing trend of LBG surface
densities as a function of redshift from $z\!\simeq\!2.0$ to $6.0$,
which we will address quantitatively in the next section.

The bottom panel of \figref{fig:ncounts} shows number counts for each
dropout sample. The \wfcnuv- and \wfcuv-dropout samples show
comparable number counts and agree generally with surveys at higher
redshifts, but the \wfcfuv-dropouts show lower number counts.  Given
the trend with redshift shown by other samples in the upper panel of
\figref{fig:ncounts}, we would expect more \wfcfuv-dropouts than other
dropouts at higher redshifts. The numbers are smaller than expected
because of the conservative selection criteria we applied owing to the
absence of a second filter below the Lyman break (see \secref{sample})
to confirm our dropout selection.  This approach led us to small
numbers of \wfcfuv-dropouts in a relatively narrow redshift range
around $z\!\simeq\!1.7$. Hence, we don't have a fully representative
\wfcfuv-dropout sample, but with the future deeper observations we can
use a somewhat more liberal selection criteria to get better
statistics for this sample.

\subsection{Determination of the UV Luminosity Function}\label{lfs}

We calculated the rest-frame UV luminosity functions (LF) using the
$V_{\rm eff}$ method \citep[e.g.,][]{stei99,sawi06,ly09} in 0.5~mag
wide bins.  The absolute magnitudes of LBG candidates
were measured in the observed bands that are equivalent to rest-frame
1500~\AA\ to minimizes $k$-corrections, and using the average redshift
for each object in each sample ($z\!\simeq\!1.7$, $2.1$, $2.7$,
respectively).  These absolute magnitudes are uncorrected for internal
dust absorption.

\begin{figure}
\includegraphics[scale=0.675,angle=90]{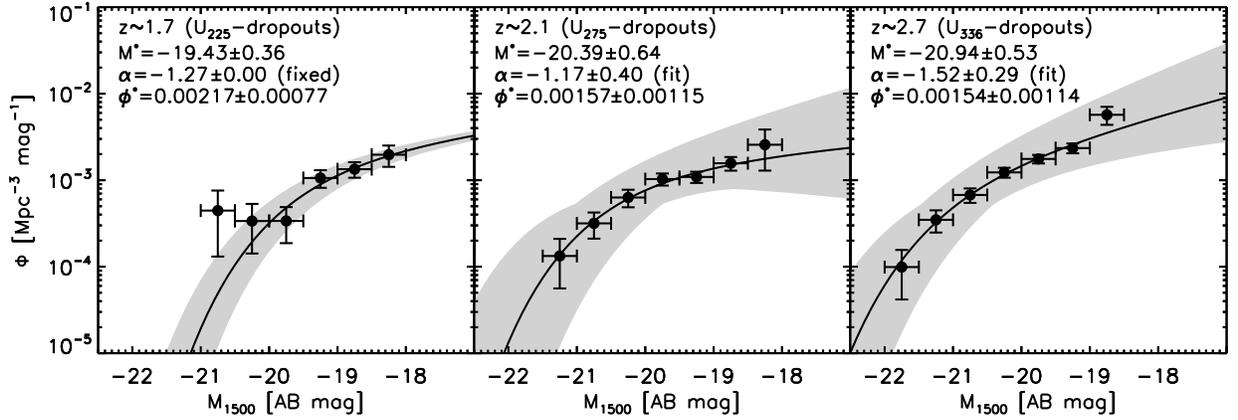}
\caption{Rest-frame UV luminosity functions for \uvdrops. The best fit
  Schechter function parameters are shown for each LF. The gray shaded
  region shows uncertainty in the LF based on 1-$\sigma$ uncertainties
  in $\alpha$ and $M^*$. In the leftmost panel, $\alpha$ is kept fixed
  while fitting the Schechter function, so the uncertainty indicated
  with the gray shaded region is based on 1-$\sigma$ uncertainty of $M^*$
  only. The more uncertain brightest point in the leftmost panel occurs because
  of very limited statistics and does not contribute to the best fit
  parameters, while the uncertain faintest points in the middle and
  the rightmost panel are at the limit of our observations, and could
  also be affected by low redshift interlopers. The vertical error
  bars in our data are 1$\sigma$ Poisson uncertainties.}\label{fig:lf}
\end{figure}

We compute LFs for the three dropout samples:
\wfcfuv-dropouts ($z\!\simeq\!1.7$ LBG candidates), \wfcnuv-dropouts
($z\!\simeq\!2.1$ LBG candidates), and \wfcuv-dropouts
($z\!\simeq\!2.7$ LBG candidates).  \figref{fig:lf} shows the LFs for
these three dropout samples. We model these LFs with a standard
Schechter function \citep{sche76}, which is parametrized by the
characteristic absolute magnitude ($M^{*}$), the normalization
($\phi^{*}$), and the faint-end slope ($\alpha$).  The shaded gray
regions in \figref{fig:lf} show the uncertainty in the LF based on
1-$\sigma$ uncertainty in $M^*$ and $\alpha$. \tabref{tab:lfs} lists
the best-fit Schechter function parameters $M^*$, $\alpha$ and
$\phi^*$ for these three dropout samples.

To investigate incompleteness in each redshift bin, we ran simulations
to calculate $P(m,z)$, which is the probability that a galaxy of
apparent magnitude $m$ and at redshift $z$ will be detected in the
image \emph{and} will meet our color selection criteria. In these
simulations, large numbers of artificial objects with a range of
redshifts and magnitudes were added to the real ERS2 images, and then
recovered using exactly the same method and selection criteria that
were employed for the real observations. For these simulations, we
used BC03 models assuming Salpeter Initial Mass Function (IMF),
constant SFR, solar metallicity, $E(B-V)=0.0-0.3$, an age of 1~Gyr with
different redshift range for each sample and varying magnitudes. These
models were used to generate color and extinction properties of our
artificial objects. We chose artificial objects to be point-like
sources. The selection function obtained from this exercise (adding
and recovering artificial objects) is similar in shape as the
distributions in \figref{fig:redshifts}, and the mean redshift value
obtained from these simulations (for each sample) is within 1-sigma of
the mean value obtained in \figref{fig:redshifts}.  These $P(m,z)$
estimates were used to determine $V_{\rm eff}$ for the LF.

We did not make the corrections for interlopers in our LF estimates.
There are five main reasons for this.  First, we have checked our LFs
by boosting the errors by 10\%, and we find that the best fit values
remain the same, while the uncertainties on these values increases
slightly.  Second, the limited number of spectroscopic redshifts
($\lesssim30$\% candidates have spectroscopic redshifts) does not give
us a correct estimate of interlopers in our color selected
sample. Third, the total fraction of spectroscopic interlopers is very
small ($\lesssim9$\%), and when we subdivide them as a function of
magnitude it is even smaller. Fourth, the estimate of interlopers
based on photometric redshifts is not very accurate because of uncertainty
in the photometric redshifts. Though the ERS2 photometric redshifts
(Cohen~et~al.~2010, in~prep) are better than some of the publicly
available photometric redshifts, they are still uncertain by a few
percent. Finally, for the \wfcnuv- and \wfcuv-dropouts the faintest
bin is most affected by the interlopers, but that data point is
already uncertain because of very few objects in that bin.

\subsubsection{Luminosity Functions}

The leftmost panel of \figref{fig:lf} shows the resulting LF for
\wfcfuv-dropouts. The three brightest bins contain on average 3
objects per bin, and hence they are more uncertain. That leaves us
with only three data points with a statistically significant number of
objects. It is not possible to fit a Schechter function to three data
points by keeping all three parameters free. In the absence of deeper
data, we fix the faint-end slope, $\alpha$, based on the best-fit
observed trend between redshifts and $\alpha$ for LBGs at
$z\!\simeq\!1.5\!-\!8$ (\figref{fig:mstr}). The best-fit parameters
for this dropout sample are meant to be mostly illustrative due to low
number statistics.  The middle panel of \figref{fig:lf} shows the LF
for the \wfcnuv-dropouts and the rightmost panel shows the LF for the
\wfcuv-dropouts.  It is difficult to estimate the faint-end slope from
these observations, as we can see from the uncertain faintest point in
the LFs of \wfcnuv- and the \wfcuv-dropouts. Though our best-fit
estimates are very close to what we expect at these redshifts
(\uvdrops) from other studies \citep[e.g.,][]{stei99,ly09} at nearby
redshifts, we will need deeper ($\sim$1--2 mag) UV observations to
properly constrain the faint-end slope for these three dropout
samples.

\subsubsection{Redshift Evolution of $M^*$ and $\alpha$}

\begin{figure}
\epsscale{0.80}
\plotone{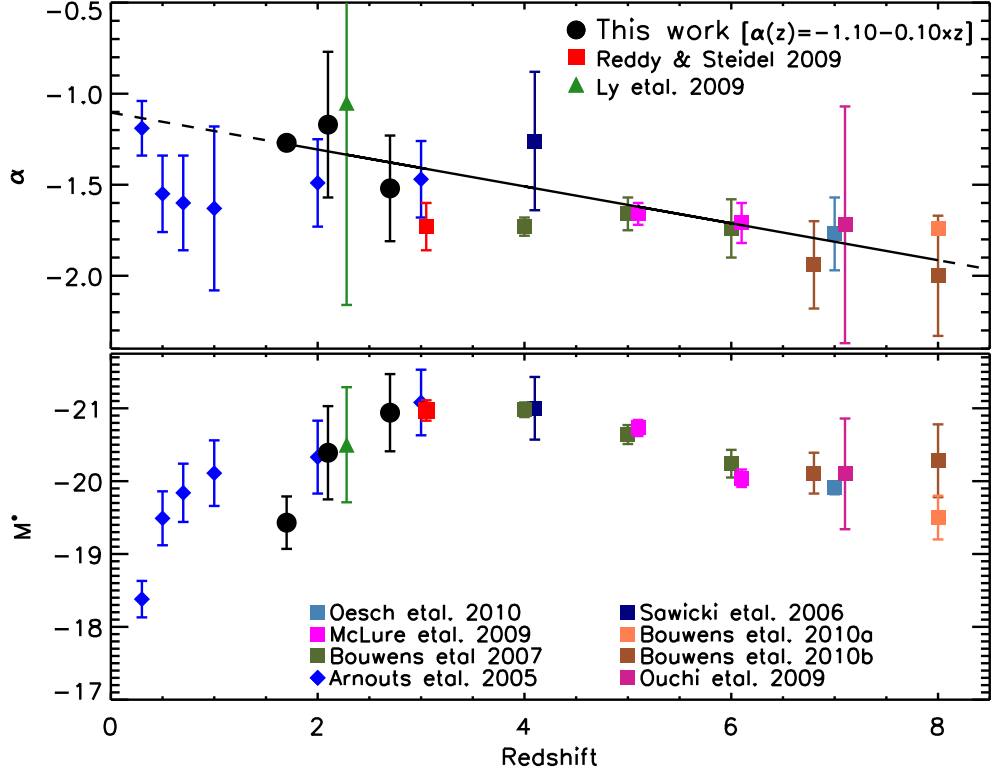}
\caption{ The best fit Schechter function parameters at rest-frame
  far-UV wavelengths --- the faint-end slope $\alpha$ (top panel) and
  the characteristic absolute magnitude $M^*$ (bottom panel) --- as a
  function of redshift. The solid black line in the top panel is the
  best fit $\alpha$--$z$ relation for LBGs at $z\!\simeq\!1.5\!-\!8$
  and extended on both sides by the dashed line. The $z\!\simeq\!1.7$
  $\alpha$ value is fixed because not enough data was available to fit
  it.  The \citet{arno05} $z\!\lesssim\!1.5$ sample is based on the
  spectroscopically confirmed galaxies with the GALEX near-UV
  detection, and the $z\!>\!1.5$ sample is based on the photometric
  redshifts.  The \citet{redd09} and the \citet{ly09} samples are
  dropout selected LBGs at $z\!\simeq\!3$ and $z\!\simeq\!2.2$,
  respectively.  The data points at $z\!\simeq\!3\!-\!8$ are for LBGs
  at high redshift.  Our results (black filled circles) agree very
  well with the general evolutionary trend observed for both these
  parameters.}\label{fig:mstr}
\end{figure}

In general, it is not straightforward to directly compare our LFs with
those from previous studies. First, our redshift range is different,
and this is the first time that this camera and filter set have been
used to select LBGs.  Secondly, in some cases the adopted cosmologies
are slightly different.  It is well known \citep[e.g.,][]{sawi06} that
the derived LFs strongly depends on the assumed cosmological models,
but the evolutionary trends seen in the LFs in our three redshift bins
are virtually independent of the assumed cosmology. \figref{fig:mstr}
shows the evolutionary trends in our three redshift bins, as well as
comparisons to other studies on LBGs at different redshifts.  

The top panel of \figref{fig:mstr} shows the faint-end LF slope,
$\alpha$, as function of redshift.  The \citet{arno05}
$z\!\lesssim\!1.5$ sample is based on the spectroscopically confirmed
galaxies with the GALEX near-UV detection ($\lesssim24.5$~mag), and
the $z\!>\!1.5$ sample is based on the photometric redshifts.  The
\citet{arno05} samples are not selected based on Lyman break color
criteria but because of the lack of LBG candidates at
$z\!\lesssim\!2.0$, we have used this star-forming galaxies sample for
comparison.  The \citet{redd09} and the \citet{ly09} samples are
dropout selected LBGs at $z\!\simeq\!3$ and $z\!\simeq\!2.2$,
respectively.  The black line is the best-fit observed trend between
$\alpha$ and $z$ for LBGs at $z\!\simeq\!1.5\!-\!8$, which is very
similar to that of \citet{ryan07}. The observed trend is that as the
redshift increases, the faint-end slope, $\alpha$, becomes steeper
(more negative), illustrating that lower luminosity dwarf galaxies
dominate the galaxy population at higher redshifts. Excluding the
fixed $\alpha$ data point at $z\!\simeq\!1.7$, our data points at
$z\!\simeq\!2.1$ and $z\!\simeq\!2.7$ agree very well --- within the
current uncertainties --- with the black line, as well as with other
data points in close redshift proximity.  

The bottom panel of \figref{fig:mstr} shows the characteristic
absolute magnitude, $M^*$, as a function of redshift. Again, the
general observed trend is that as redshift increases, the
characteristic absolute magnitude, $M^*$, becomes brighter (more
negative) until $z\!\simeq\!3.5$.  This trend is considered as an
evidence of `downsizing' galaxy formation scenario
\citep[e.g.,][]{cowi96}, where luminous massive galaxies form at
higher redshifts. Our first data point at $z\!\simeq\!1.7$ follows
this general trend, but it is more uncertain due to the limited
statistics in this dropout sample. The other two data points at
$z\!\simeq\!2.1$ and $2.7$ fit very well within the evolutionary trend
seen at these and higher redshifts.  \figref{fig:mstr} shows rapid
decline of $M^*$ between $z\!\simeq\!3$ and extending to
$z\!\simeq\!1.5$. This turnover is well defined in our and
\citet{ly09} samples. It is important for future surveys to exploit
the special capabilities of the WFC3 in the near-UV to obtain larger
samples to understand the relation between this critical transition in
$M^*$ and physical processes in LBGs at $z\!<\!3$.

\citet{redd09} have used deep ground-based imaging data to
constrain the UV LF of the `BX' \citep[e.g.,][]{adel04} population at
$1.9\!<\!z\!<\!2.7$, which selects star-forming galaxies based on
$U_{n}GR$ colors. When we compare our LFs with that of the `BX'
population, we find some differences in $M^*$ and $\alpha$ values. First,
our \wfcfuv-dropout sample has lower redshift ($z\!\sim\!1.7$)
compared to the `BX' population ($z\!\sim\!2.3$), so our $M^*$ value
(--19.43~mag) is fainter than their value of --20.70~mag, and agrees
with the general trend discussed above. Second, $M^*$ values of our
\wfcnuv-dropout sample and the `BX' sample agree within our 1-$\sigma$
uncertainty, while the $\alpha$ is little steeper for the `BX' sample.  We
believe that complete agreement between our LBGs sample and the `BX'
sample is not possible, because although the `BX' selection selects
star-forming galaxies, it is very likely that the dropout selected
sample at a similar redshift might not be same as the `BX' selected
sample.  Some galaxies which are selected through the `BX' color
selection criteria might not be in the dropout selected sample, and
vice versa. Therefore, it is difficult to directly compare the `BX'
and the LBG samples, and the differences in these samples could cause
the LF parameters at similar redshift to differ \citep[see also][]{ly09}. 

Therefore, for both $M^*$ and $\alpha$ our results agree very well
with the expected observed trends (\figref{fig:mstr}) as a function of
redshift. At lower redshifts ($z\!<\!3$), our data points will help to
reduce the gap between the well studied $z\!\gtrsim\!3$ and
$z\!\sim\!0$ regimes.  The agreement with the observed evolutionary
trend of M$^*$ and $\alpha$ also show the reliability of our LFs,
which can be improved with the future deeper and wider WFC3 UV
observations (e.g., CANDELS Multi-Cycle Treasury program \#
12060-12064, PI: S.~Faber).

%--------------------------------------------------

\section{Summary}\label{conclusion}

We use newly acquired UV observations from the WFC3 UVIS channel along
with existing deep ACS observations of the GOODS program, to identify
UV dropout galaxies, which are LBG candidates at \uvdrops. We find 66
\wfcfuv-, 151 \wfcnuv- and 256 \wfcuv-dropouts to a magnitude limit of
AB$\simeq$26.5~mag. This allows us to estimate rest-frame UV LFs in
three redshift bins ($z\!\simeq\!1.7$, $2.1$, $2.7$). Their best-fit
Schechter function parameters $M^{*}$, $\alpha$ and $\phi^{*}$ agree
very well with the observed evolution of these parameters with respect
to redshift. We need space-based UV imaging to identify and understand
the $z\!\lesssim\!3$ LBGs selected based on their UV dropout
signature. The new WFC3 UVIS camera on the HST now allows us to do
that with much better sensitivity and resolution than GALEX, and has
opened up a new regime of detailed UV imaging studies of low to
intermediate redshift ($z\!\lesssim\!3$) LBGs, which is not possible
from the ground due to the atmospheric cut-off. The quality of
rest-frame near-UV imaging ($\gtrsim$3000~\AA) of these galaxies
greatly exceeds that which can be done with ground-based near-UV
observations.  Future work will investigate the morphology and stellar
populations of these lower redshift LBGs, to better understand their
higher redshift counterparts.  The upcoming WFC3 UVIS imaging surveys
--- deep and wide --- have the potential to robustly measure the
evolution of LBGs at $z\!\lesssim\!3$, and --- with uniform selection
all-the-way to very high redshifts --- provide better understanding of
very high redshift LBGs.

%--------------------------------------------------
\acknowledgments

We thank the referee for helpful comments and suggestions that
significantly improved this paper.  We thank M.~Nonino, C.~Ly and
their collaborators for providing their LBG number counts.  This paper
is based on Early Release Science observations made by the WFC3
Scientific Oversight Committee. We are grateful to the Director of the
Space Telescope Science Institute for awarding Director's
Discretionary time for this program. Finally, we are deeply indebted
to the brave astronauts of STS-125 for rejuvenating HST. Support for
program \#11359 was provided by NASA through a grant from the Space
Telescope Science Institute, which is operated by the Association of
Universities for Research in Astronomy, Inc., under NASA contract NAS
5-26555.

%--------------------------------------------------

%--------------------------------------------------
% Figure 1 -- lyman breaks and filter curves

\clearpage

%--------------------------------------------------
% Figure 2 -- color-color selection plots

\clearpage

%--------------------------------------------------
% Figure 3 -- example images of 3 dropouts

\clearpage

%--------------------------------------------------
% Figure 4 -- redshift comparisons

\clearpage

%--------------------------------------------------
% Figure 5 -- number counts

\clearpage

%--------------------------------------------------
% Figure 6 -- luminosity functions

\clearpage

%--------------------------------------------------
% Figure 7 -- redshift evolution of the LF

\clearpage

%--------------------------------------------------
% Table 1

\begin{deluxetable}{cccc}
\tablewidth{0pt}
\tablecaption{Number and Redshifts of UV-dropouts in the ERS2 field \label{tab:redshifts}}
\tablenum{1}

\tablehead{\colhead{Dropout} & \colhead{Total Number} & \colhead{No. of Spectroscopic} & \colhead{No. of Photometric} \\
  \colhead{Filter} &      \colhead{of Candidates} &   \colhead{Redshifts$^{a}$} &      \colhead{Redshifts$^{b}$} \\
  \colhead{} & \colhead{} & \colhead{and $<\!z\!>$$^{c}$} &
  \colhead{and $<\!z\!>$$^{c}$}}

\startdata
$U_{\rm 225}$ & 66 & 22                      & 55    \\
        &    & 1.59$\pm$0.22 & 1.51$\pm$0.13 \\
$U_{\rm 275}$ & 151 & 57                      & 117   \\
        &     & 2.07$\pm$0.40 & 2.09$\pm$0.42 \\
$U_{\rm 336}$ & 256 & 52                      & 203   \\
       &     & 2.40$\pm$0.40 & 2.28$\pm$0.40 \\
\enddata

\tablenotetext{a}{From compilation of VLT redshifts \citep[e.g.,][]{graz06,wuyt08,vanz08,bale10}}
\tablenotetext{b}{From Cohen~et~al.~2010, in prep.}
\tablenotetext{c}{From the Gaussian fit to the distribution shown in \figref{fig:redshifts}}
\end{deluxetable}

%--------------------------------------------------
% Table 2

\begin{deluxetable}{cccc}
\tablewidth{0pt}
\tablecaption{Parameters of Schechter Function Fits \label{tab:lfs}}
\tablenum{2}

\tablehead{\colhead{Dropout} & \colhead{$M^*$ (1500~\AA)}  & \colhead{$\phi^*$ } & \colhead{$\alpha$} \\
  \colhead{Filter} & \colhead{(AB mag)} & \colhead{mag$^{-1}$
    Mpc$^{-3}$} & \colhead{} }
	
\startdata
$U_{\rm 225}$ & --19.43$\pm$0.36 & 0.00217$\pm$0.00077  & --1.27 (fixed) \\
$U_{\rm 275}$ & --20.39$\pm$0.64 & 0.00157$\pm$0.00115 & --1.17$\pm$0.40  \\
$U_{\rm 336}$ & --20.94$\pm$0.53 & 0.00154$\pm$0.00114   & --1.52$\pm$0.29  \\
\enddata

\end{deluxetable}

%--------------------------------------------------

\end{document}